# Two ways for numerical solution of the Kramers problem for spatial diffusion over an edge-shaped barrier

**M V Chushnyakova[1], I I Gontchar[2], A. V. Zakharov[3], N. A. Khmyrova[2]**

[1] Physics Department, Omsk State Technical University, Omsk 644050, Russia
[2] Physics and Chemistry Department, Omsk State Transport University, Omsk 644046, Russia
[3] Pre-university Training Department, Omsk State Technical University, Omsk 644050, Russia

maria.chushnyakova@gmail.com

**Abstract**. Thermal decay rate over an edge-shaped barrier at high dissipation is studied numerically through the computer modeling. Two sorts of the stochastic Langevin type equations are applied: (i) the Langevin equations for the coordinate and conjugate momentum (LEqp, the phase space diffusion) and (ii) the reduced Langevin equation (RLE, the spatial diffusion, overdamped motion). The latter method is much faster and self-similar; however, one can doubt about its applicability in the case of an edge-shaped barrier with a discontinuous force. The reason is that a formal condition of the applicability of the RLE is not fulfilled since the curvature of the potential profile at the barrier is equal to infinity. The present numerical study demonstrates that, for large friction, the decay rate calculated using the RLE agrees with the rate resulting from the more exact LEqp. Moreover, it turns out that the influence of the position of the absorbing border is similar to the case of harmonic potential known in the literature.

## 1. Introduction

Thermal decay of a metastable state (i.e. diffusion of Brownian particles from a potential well due to thermal fluctuations) is a useful model in natural sciences [1,2]. One of the regimes, the so-called overdamped motion (spatial diffusion), is particularly useful in biophysics where the nanomotors propulsion occurs in the viscous water medium [3,4]. For this regime, several approximate analytical formulas were derived by Kramers almost eighty years ago [5]. Accuracy of some of these formulas has been tested numerically [6,7] and the corrections have been obtained [8] which improve the agreement of the analytical rates with the numerical ones.

However, one case considered by Kramers has not found its numerical verification so far, the case of an edge-shaped barrier formed by two parabolas crossing each other (see Fig. 1). In the present work, we are going to model numerically diffusion of Brownian particles from the left pocket to the right one considering the well-known first passage time problem [1,2,9]. The decay rate is the main characteristic of this process. We operate with two dimensionless parameters [10] defining mostly the value of the rate: a governing parameter $G$ and a damping parameter $\varphi$:

$$G = \frac{U_b}{\theta}, \qquad (1)$$

$$\varphi = \frac{\eta \tau_c}{2\pi m}. \tag{2}$$

Here $U_b$ is the height of the potential barrier; $\theta$ is the average thermal energy; $\tau_c$ is the oscillation period of the Brownian particle near the bottom of the potential well; $m$ and $\eta$ are the inertia and friction parameters, respectively. Those are analogous to the mass of the Brownian particle and the resistance coefficient of the medium.

One finds in the literature two ways for numerical modeling of the decay process. First, the Langevin equations for the coordinate and conjugate momentum (LEqp) are used [2,11]; this approach corresponds to the phase space diffusion. This is the most precise way for modeling; however, it is very computer time consuming. In the second approach, which is valid for the so-called spatial diffusion regime, a reduced Langevin equation (RLE) is applied [2,6]. The RLE is approximate but very attractive because it is self-similar with respect to the friction parameter and the corresponding computer modeling is much faster.

On the one hand, the condition of applicability of the RLE is that the equilibrium momentum distribution is established significantly faster than the coordinate distribution evolves. On the other hand, there is a formal condition for the rate to be evaluated by the overdamped Kramers formula. The condition reads

$$\frac{\eta}{m} \gg \omega_b. \tag{3}$$

Here $\omega_b$ is the frequency-like quantity characterizing the motion around the barrier; for the case of the edge-shaped barrier shown in Fig 1 this "frequency" is infinite. This circumstance places some doubt on the applicability of the approximation of the spatial diffusion in the case under consideration.

The purpose of our work is to check whether the approximate description with the RLE is valid in the case of the edge-shaped barrier.

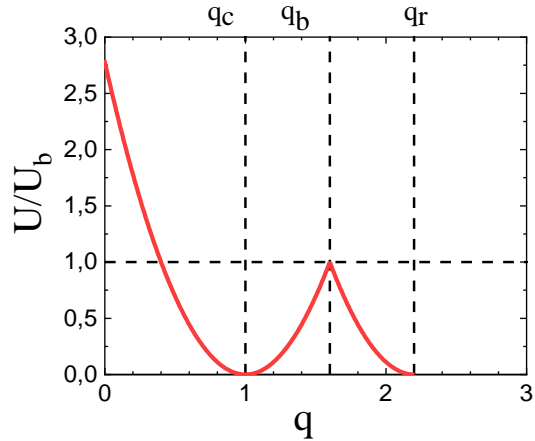

**Figure 1.** The edge-shaped potential in which a Brownian particle wanders starting from $q_c$. The coordinates of the barrier ($q_b$) and of the bottom of the right parabola ($q_r$) are also shown.

## 2. Numerical model

We use two approaches based on the stochastic Langevin-type equations. In the first approach, the equations for the phase space diffusion (we abbreviate those as LEqp) are used:

$$\frac{dq}{dt} = \frac{p}{m}, \tag{4}$$

$$\frac{dp}{dt} = -\frac{\eta}{m}p - \frac{dU}{dq} + \sqrt{\eta\theta}\,\Gamma(t). \tag{5}$$

Here $q$ and $p$ stand for the generalized coordinate and its conjugate momentum, respectively; the white noise $\Gamma(t)$ possesses the following statistical properties: $\langle \Gamma(t) \rangle = 0$, $\langle \Gamma(t_1)\Gamma(t_2) \rangle = 2\delta(t_1 - t_2)$.

An advantage of the LEqp is that these equations are exact, i.e. other equations describing the Markovian process with Brownian particles in more detail are not known. However, these equations are very time consuming in computer modeling. In particular, increasing friction we must decrease the time step of the modeling which makes the modeling slower and slower.

The second approach is based on a single equation (the reduced Langevin equation, RLE):

$$\frac{dq}{dt} = -\frac{1}{\eta}\frac{dU}{dq} + \sqrt{\frac{\theta}{\eta}}\,\Gamma(t). \tag{6}$$

The computer modeling by means of this equation is much faster than by means of the LEqp. Moreover, as the friction increases, one can use a larger time step.

Another advantage of the RLE is that it is self-similar: after finding $R_{at}(t, \eta_1)$ one easily obtains $R_{at}(t, \eta_2)$ as follows

$$\eta_2 R_{at}(t/\eta_2, \eta_2) = \eta_1 R_{at}(t/\eta_1, \eta_1). \tag{7}$$

However, Eq. (6) produces a meaningful result only in the overdamping regime, i.e. at some stage, the validity of its result must be checked using the LEqp.

Solving the Langevin equations (LEqp or RLE) one obtains $N_{tot}$ trajectories, each of those ends up not later than at $t_D$. Some trajectories (useful trajectories) reach the absorbing border $q_a$ at $t_a < t_D$. The time-dependent decay rate, based on these trajectories, reads

$$R_{at}(t) = \frac{1}{N_{tot} - N_{at}} \frac{\Delta N_{at}}{\Delta t}. \tag{8}$$

Here $N_{at}$ stands for the number of trajectories reaching $q_a$ by the time moment $t$; $\Delta N_{at}$ is the number of trajectories reaching $q_a$ during the time lapse $\Delta t$. The time dependencies $R_{at}(t)$ can be found in many papers (see, e.g., [6,7,12]) for different potential shapes and/or different values of the damping parameter $\varphi$. After some delay time, the rate reaches a quasistationary value $R_D$. This rate is time independent within the statistical fluctuations. For calculating $R_D$, we use several bins starting from the tail of the $R_{at}(t)$- array and average over these bins. In the present work, the number of modeled trajectories and $t_D$ are chosen to obtain the value of $R_D$ with the relative error not exceeding 2%. The computer code used in the present work was verified numerously [7,10,12].

## 3. Approximate Kramers rates

In addition to the numerical algorithm, it is always desirable to have an analytical formula to estimate any physical quantity we are interested in. Such approximate formulas were derived by Kramers in his seminal work [5]. Kramers introduced a flux-over-population method which implies that the rate $R$ is equal to the flux over the barrier, $j$, divided by the population $n$ of the potential well.

Let us shortly repeat the derivation. This is useful, in particular, because Kramers himself did not write the final formula explicitly. The derivation starts from the Smoluchowski equation for the probability density $\sigma(q, t)$

$$\frac{\partial \sigma(q,t)}{\partial t} = -\frac{\partial j(q,t)}{\partial q} = \frac{\partial}{\partial q}\left(\frac{\sigma}{\eta}\frac{dU}{dq} + \frac{\theta}{\eta}\frac{\partial \sigma}{\partial q}\right). \tag{9}$$

In the stationary regime, it is convenient to transform the flux as follows

$$j = -\frac{\theta E}{\eta}\frac{d}{dq}\left(\frac{\sigma}{E}\right) \tag{10}$$

where $E = \exp(-U(q)/\theta)$ is proportional to the equilibrium distribution. Then Kramers transformed Eq. (10) to the following form

$$\frac{j\eta}{\theta E} = -\frac{d}{dq}\left(\frac{\sigma}{E}\right) \qquad (11)$$

and integrated both sides of Eq. (11) over the coordinate from $q_c$ up to $q_a$. Considering the flux $j$ to be both coordinate and time independent, Kramers took $j$ out of the integral. These transformations result in

$$j = \frac{\theta}{\eta}\left.\frac{\sigma}{E}\right|_{q_a}^{q_c}\left(\int_{q_c}^{q_a}\frac{1}{E(q)}dq\right)^{-1}. \qquad (12)$$

The population of the well reads

$$n = \int_{-\infty}^{q_b}\sigma(q_c)E(q)dq. \qquad (13)$$

The quasistationary decay implies that almost all particles are located near the bottom of the left well, i.e. $\sigma(q_c) \gg \sigma(q_a)$. Suggesting that the equilibrium distribution in the left well is established and substituting Eqs. (12) and (13) into the flux-over-population formula, one obtains

$$R_{IK} = \frac{\theta}{\eta}\left\{\int_{-\infty}^{q_b}\exp\left[-\frac{U(y)}{\theta}\right]dy \int_{q_c}^{q_a}\exp\left[\frac{U(z)}{\theta}\right]dz\right\}^{-1}. \qquad (14)$$

Equation (14) was not written explicitly in the Kramers paper, maybe because, at that time, there were no computers to evaluate the integrals. However, from the text of [5], it is absolutely clear that Kramers meant this formula. Ironically, it took seventy years for this formula to be written explicitly: by our knowledge, it was published for the first time in [6] in 2010. Let us call $R_{IK}$ the integral Kramers rate.

Due to the absence of computers, Kramers proceeded making approximate estimations for the integrals. For the first integral in Eq. (14) he extended the upper limit up to $+\infty$ reducing the integral to the Poisson's one. The second integral for the symmetric edge-shaped barrier of Fig. 1 reads

$$\int_{q_c}^{q_a}\exp\left(\frac{U(z)}{\theta}\right)dz = 2\int_{q_c}^{q_b}\exp\left(\frac{U(z)}{\theta}\right)dz. \qquad (15)$$

The latter integral is evaluated by the Laplace method since the integrand reaches its maximum value at $z = q_b$:

$$\int_{q_c}^{q_b}\exp\left(\frac{U(z)}{\theta}\right)dz = \exp\left(\frac{U(q_b)}{\theta}\right)\frac{\theta}{U'(q_b)}. \qquad (16)$$

These transformations result in

$$R_K = \frac{\sqrt{\pi G}}{\varphi \tau_c} \cdot \exp(-G). \qquad (17)$$

As in the case of the barrier composed by two smoothly joined parabolas (harmonic barrier) [10], this rate (measured in units of $\omega_c = 2\pi/\tau_c$) depends only upon the dimensionless scaling parameters $\varphi$ and $G$ (see Eqs. (1) and (2) for the definitions of these parameters).

### 4. Results
In Fig. 2 we compare the dynamical time-dependent rates obtained by means of the LEqp, $R_{atqp}$ (wriggling curves with open symbols), and the rates resulting from the RLE, $R_{atq}$ (wriggling curves with solid symbols). Note, that obtaining the former rate requires by a factor of 20 longer computations. Thin horizontal lines indicate the quasistationary values of the rates (dash-dotted lines for the LEqp and dashed lines for the RLE). The Kramers approximate rates evaluated through Eq. (17) are shown in Fig. 2 as well (thick solid horizontal lines). First of all, one sees that the full modeling of the process in the phase space using the LEqp and the simplified modeling by means of the RLE result in the very same time-dependent rate at $\varphi$ =15.3, and, consequently, for the higher values of this parameter too: the quasistationary values are very close as well as the transient stages. This is a significant advantage because the RLE is less time consuming.

The second point which is seen from Fig. 2 is that the Kramers rate $R_K$ is only in qualitative agreement with the dynamical quasistationary rate: the difference is about 30%.

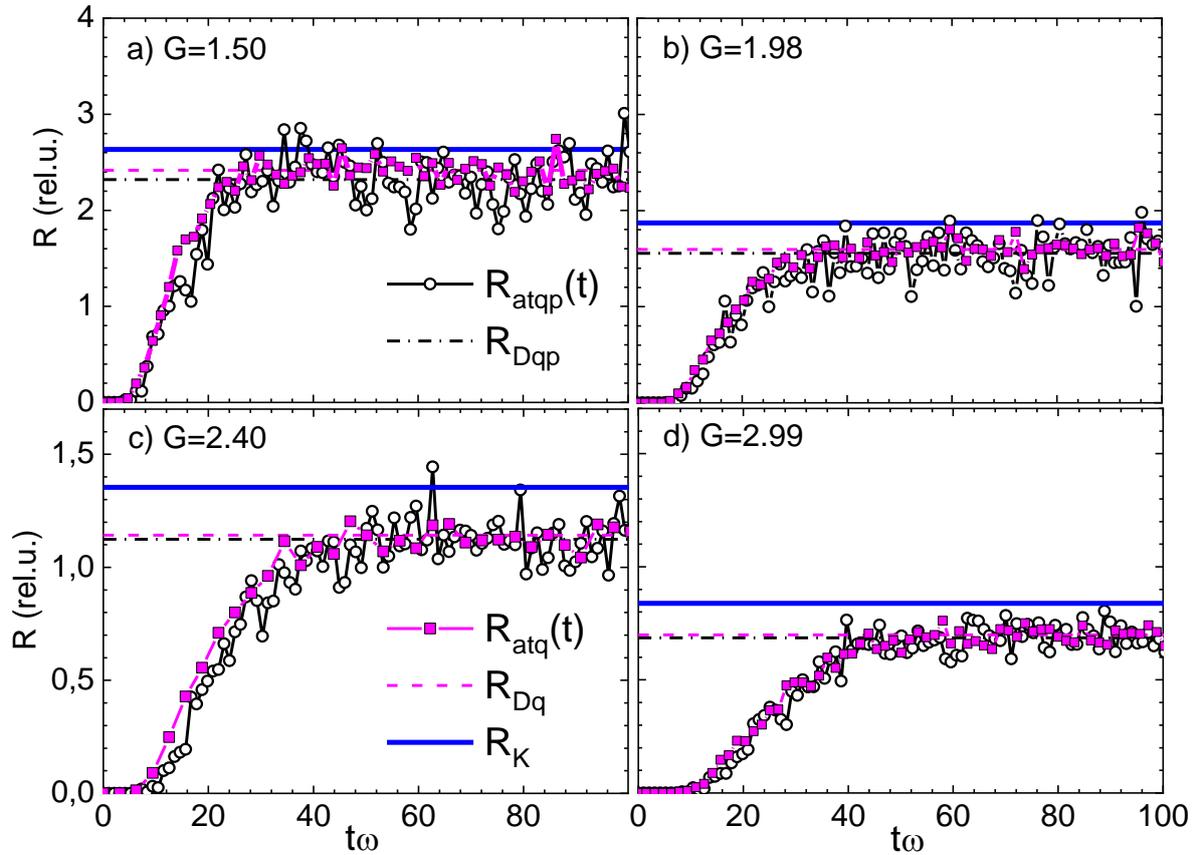

**Figure 2.** The rates $R_{atqp}$ and $R_{atq}$ (wriggling curves with symbols) are shown as the functions of time for four values of the governing parameter. The corresponding quasistationary rates and the approximate rate $R_K$ (horizontal lines) are presented as well. $\varphi$=15.3.

In Fig. 3 we again compare the rates resulting from the two sorts of dynamical modeling. In fact, it was never proved before that the approach based on the reduced Langevin equation produces the results in agreement with the exact LEqp for the edge-shaped barrier. One could doubt about such agreement considering the well-known Kramers formula for medium and large friction the decay rate over the harmonic barrier formed by two smoothly joint parabolas of different stiffnesses:

$$R_{KP} = \frac{\omega_c}{2\pi\omega_b}\left\{\left(\frac{\beta^2}{4}+\omega_b^2\right)^{1/2}-\frac{\beta}{2}\right\}\cdot \exp(-G). \tag{18}$$

Here $\beta = \eta/m$ is the damping coefficient, $\omega_c$ ($\omega_b$) is the frequency corresponding to the potential minimum (barrier). When the formal condition (3) is fulfilled, Eq. (18) is reduced to the formula analogous to Eq. (17). Obviously, for the edge-shaped barrier $\omega_b$ is infinite and condition (3) is not obeyed.

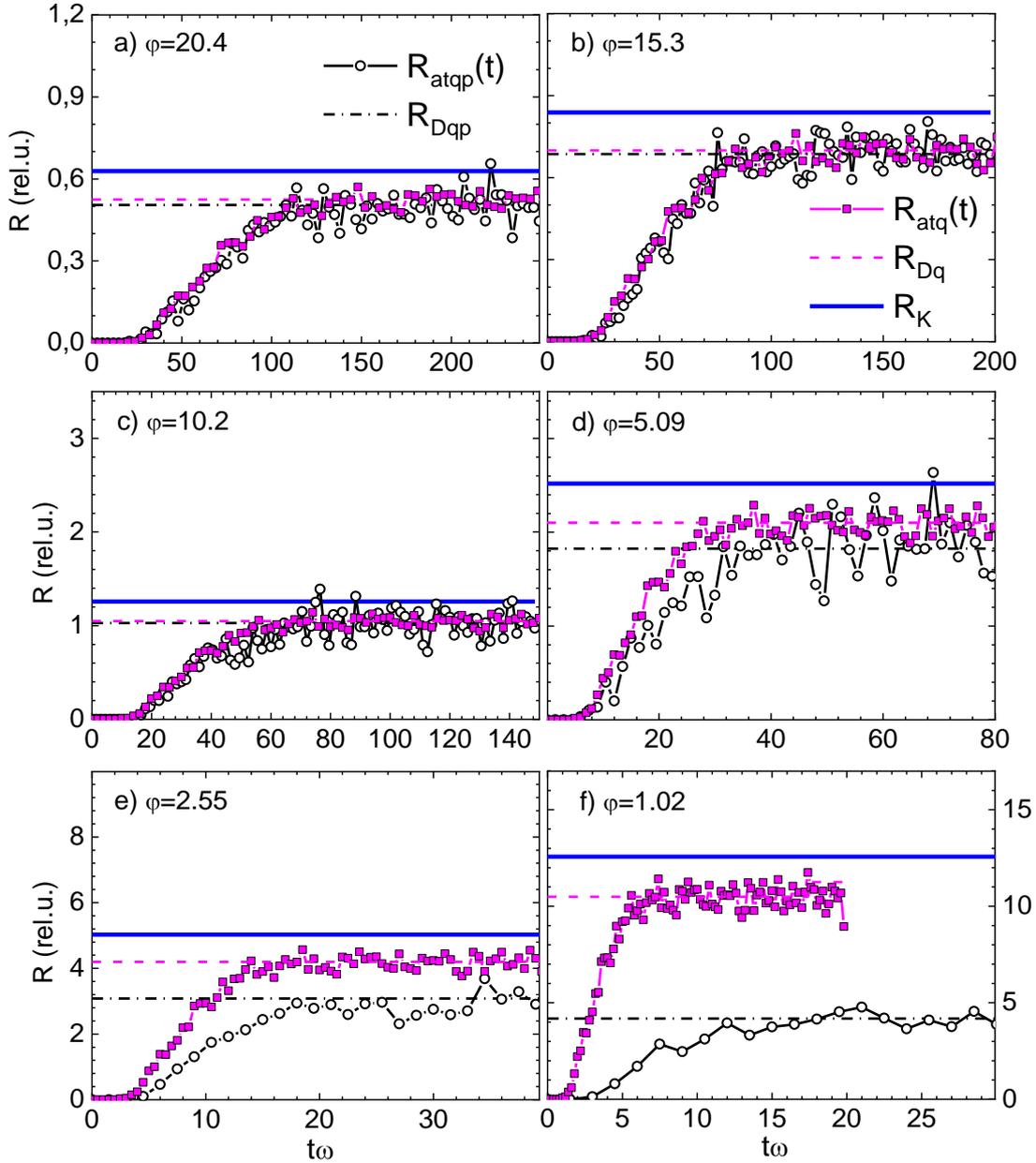

**Figure 3.** Same as in Fig. 2 but for six values of the damping parameter. $G=3.0$

Nevertheless, we see in Fig. 3 that up to $\varphi = 10$ the rate $R_{atq}$ is in good agreement with $R_{atqp}$. However, at the damping parameter equal to 5 the difference becomes visible reaching a factor of 3 at $\varphi = 1$. Note, that the rates $R_{atq}$ used in this figure are obtained on the basis of the only numerical rate $R_{atq}(\varphi = 15)$ by scaling it according to Eq. (7). This saves a lot of computer time.

It is known from the former studies for the harmonic barrier [2,7,13] that, as the absorptive border moves closer to the barrier, the quasistationary rate becomes twice the rate at the remote $q_a$. It is interesting to check whether this result survives for the edge-shaped barrier of Fig. 1. For this goal, we perform calculations for three positions of the absorptive border: at the bottom of the right parabola ($q_a = q_r$, see Fig. 1), at the barrier ($q_a = q_b$), and at the intermediate point ($q_a = 2.0$). Note, that all the previous calculations are performed using $q_a = q_r$. The results of this study are illustrated by Fig. 4.

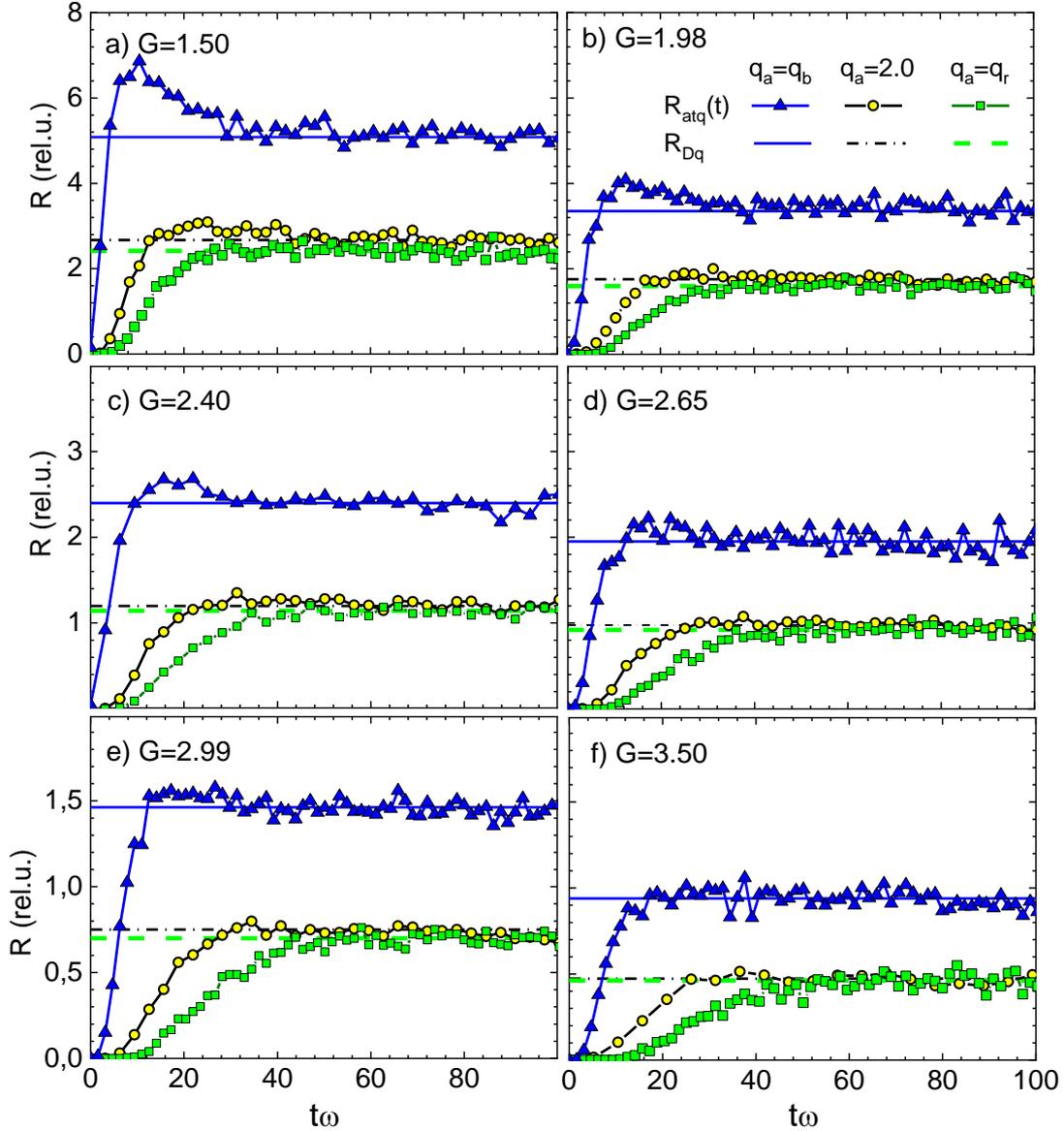

**Figure 4.** The rates $R_{atq}$ calculated at three locations of the absorbing border: triangles - $q_a = q_b$, circles - $q_a = 2.0$, squares - $q_a = q_r$ for six values of the governing parameter. The corresponding quasistationary rates are shown by the horizontal lines. $\varphi = 15.3$.

One sees that as the absorptive border moves closer to the barrier, the quasistationary dynamical rate increases similar to the case of the harmonic potential (see Fig. 2 in [14]). The ratio $R_D(q_a = q_b)/R_D(q_a = q_r)$ is indeed close to 2 for all values of the governing parameter in its range considered. This result is somewhat unexpected keeping in mind the findings of [7]. In that work, it was shown that the ratio $R_D(q_a = q_b)/R_D(q_a = q_r)$ depends upon the shape of the descent from the barrier. We are inclined to think that the ratio takes the same value for the edge-shaped and harmonic barriers because for both potentials the profiles of the ascent and descent are identical. Probably this question needs more attention, in particular, for the case of the asymmetric edge-shaped barrier.

## 5. Conclusions

In the present work, we have studied numerically the process of thermal decay rate over a symmetric edge-shaped barrier at strong dissipation. For this aim, two kinds of the stochastic differential equations have been used. First, the Langevin equations for the coordinate and conjugate momentum (LEqp) have been employed. This approach is more precise yet more computer time consuming. Second, the process has been modeled by means of the reduced Langevin equation (RLE). This approach corresponds to the spatial diffusion, i.e. to the overdamped motion; it is significantly faster and self-similar with respect to the damping parameter. However, one can doubt about its applicability in the case of the edge-shaped barrier due to the discontinuity of the force at the barrier point. We have shown that, for large friction, the decay rate calculated using the RLE is in good accord with the rate resulting from the more exact LEqp. Moreover, it has been demonstrated that the position of the absorbing border affects the value of the quasistationary rate in the manner similar to that for the harmonic barrier. This fact demands further investigation for the case of the edge-shaped potential formed by the parabolas with different stiffnesses.